# Teaching Well-Structured Code: A Literature Review of Instructional Approaches

Sara Nurollahian, Hieke Keuning, Eliane Wiese

# Teaching Well-structured Code: A Literature Review of Instructional Approaches

Sara Nurollahian
Kahlert School of Computing
University of Utah
Salt Lake City, USA
sara.nurollahian@utah.edu

Hieke Keuning
Information and Computing Sciences
Utrecht University
Utrecht, The Netherlands
h.w.keuning@uu.nl

Eliane Wiese
Kahlert School of Computing
University of Utah
Salt Lake City, USA
eliane.wiese@utah.edu

*Abstract*—Teaching the software engineers of the future to write high-quality code with good style and structure is important. This systematic literature review identifies existing instructional approaches, their objectives, and the strategies used for measuring their effectiveness. Building on an existing mapping study of code quality in education, we identified 53 papers on code structure instruction. We classified these studies into three categories: (1) studies focused on developing or evaluating automated tools and their usage (e.g., code analyzers, tutors, and refactoring tools), (2) studies discussing other instructional materials, such as learning resources (e.g., refactoring lessons and activities), rubrics, and catalogs of violations, and (3) studies discussing how to integrate code structure into the curriculum through a holistic approach to course design to support code quality. While most approaches use analyzers that point students to problems in their code, incorporating these tools into classrooms is not straightforward. Combined with further research on code structure instruction in the classroom, we call for more studies on effectiveness. Over 40% of instructional studies had no evaluation. Many studies show promise for their interventions by demonstrating improvement in student performance (e.g., reduced violations in student code when using the intervention compared with code that was written without access to the intervention). These interventions warrant further investigation on learning, to see how students apply their knowledge after the instructional supports are removed.

## I. INTRODUCTION

Students must recognize that it is not sufficient to write code that generates the correct output. The code must also be of high quality—readable, maintainable, and understandable by others. High-quality code simplifies debugging, modification, and extension, making it indispensable for both collaborative projects and long-term maintenance.

A crucial aspect of code quality that is also important for students' learning of programming is *code structure*, also referred to as *semantic style* [1]. Code structure pertains to implementation choices within a single function (e.g., the selection of control flow, assignment, and operations). These choices can be made by looking at code segments in isolation (e.g., avoiding duplicated code within conditional branches). This aspect of code quality is distinct from aspects that do not affect code execution, such as comments [2], formatting [2], and naming conventions [3]. We refer to idiomatic code with proper structure as *well-structured code*. In contrast, alternative structures, which are often verbose, are referred to as *code structure defects* or *code structure violations*. Table I shows three code structure topics, each with a violation and its well-structured counterpart. Both violations and well-structured code have the same input/output behavior for any topic, but well-structured code is simpler and more idiomatic [4], [5].

Teaching code structure to students is difficult. Students do not automatically learn to write well-structured code as they gain more programming knowledge [6]. Although there are established guidelines and standards for professional programmers to write well-structured code, such as *Clean Code* [7] and *Refactoring* [8], these may not be suitable for novice programmers. Similarly, professional code analyzers like PMD [9] and CheckStyle [10] can flag code structure violations, but these tools are not designed for beginner programmers. They often raise flags irrelevant to student learning, and their minimal feedback may be difficult for students to use [11].

Prior studies indicated that instructors, particularly those with less experience, often lack a clear understanding of important code structure topics for student learning and the types of support students need [12]. Catalogs of code structure violations (such as those by Effenberger et al. [13] and Řecháčková et al. [14]) help illuminate important topics. However, the question of how to effectively teach these topics to students remains unanswered. Although educational researchers have begun developing and exploring tools and resources for teaching code structure, we still lack a comprehensive overview of the instructional approaches being used and their effectiveness.

Thus, we conducted a systematic literature review to identify the instructional approaches used in prior studies to support students in learning about code structure. Specifically, we aim to answer the following questions:

RQ1 What instructional approaches for teaching code structure have been studied, and what goals do they aim to achieve?
RQ2 How did researchers examine the effectiveness of the approaches, and how effective were they?

For RQ1, we found that 24 studies discussed an automated *tool*, including code analyzers, tutors, and refactoring tools. Code analyzers were the most common type of tool for code structure instruction. Such tools can identify code structure violations in student programs and provide feedback to students

TABLE I: Three code structure topics related to conditionals. In each topic, both violation and well-structured code indicate the same functionality, but professional programmers prefer the latter as it is simpler, more idiomatic, and easier to modify.

| Topic | Code structure violation | Well-structured code |
| --- | --- | --- |
| Return boolean expressions instead of using `if` to return boolean literals | ```java
if (word.startsWith("A")) {
    return true;
}
return false;
``` | ```java
return word.startsWith("A");
``` |
| Modify sequential `if` statements to an `if` and `else` | ```java
if (number % 2 == 0) {
    return "Even number!";
}
if (number % 2 != 0) {
    return "Odd number!";
}
``` | ```java
if (number % 2 == 0) {
    return "Even number!";
} else {
    return "Odd number!";
}
``` |
| Simplify nested `if` statements by conjoining conditions | ```java
if (age >= 18) {
    if (hasDriverLicense) {
        return "Can drive!";
    }
}
return "Can not drive!";
``` | ```java
if (age >= 18 && hasDriverLicense) {
    return "Can drive!";
}
return "Can not drive!";
``` |

(e.g., [15], [16]), and may even enhance students' motivation and awareness about code quality (e.g., [17]).

In addition, 11 studies in our dataset discussed *teaching materials* other than tools for teaching code structure. This approach involved providing learning resources such as lessons, activities, or exercises for students to practice refactoring [18]–[20], and designing rubrics that instructors can use to provide feedback and grades on students' work [21], [22]. Only two studies in our dataset discussed their experience on *integrating code structure into the curriculum*. In the remaining studies, we identified a tool or resource that can be *potentially useful for educational purposes*, even though the study did not detail its instructional application. For example, some studies developed or used tools to assess violations in student programs [11], [23] or created catalogs of defects [13] that could support instruction.

For RQ2, we found that more than one-third of the studies that discussed an instructional approach did not assess its effectiveness with students. Among the studies that assessed the effectiveness, the most common approach was comparing the frequency of violations in student code between those who used the intervention and those who did not (e.g., [24], [25]). Other methods included asking students to self-report the tool's perceived usefulness (e.g., [15], [26]), analyzing the log data to study student behavior when interacting with the tool [26], [27], and evaluating student learning [18].

## II. BACKGROUND AND PRIOR WORK

Writing high-quality code is essential for students' success as software developers, and for creating reliable and efficient software systems. Stegeman et al. [21] define code quality as "an aspect of software quality that concerns directly observable properties of source code." This definition is extended by Kirk et al. [28], emphasizing that code quality is "constrained to understanding and changing code." Therefore, writing high-quality code encompasses more than generating the correct output; it ensures code readability, which facilitates collaboration and reduces the costs associated with maintenance.

### A. Code Structure Is An Element of Code Quality

Code quality involves many aspects, such as adhering to consistent formatting [29], using descriptive names for variables and methods [3], applying effective design principles [30], and even the approach to solving problems [31] can impact the quality. In this study, we focus on another important aspect we refer to as *code structure*. Code structure relates to the stylistic choices for implementing code within a single function, such as the use of control flow structures, assignments, and comparisons. We use the term "stylistic" to emphasize that different implementation choices do not change the code's semantic (IO) behavior. However, it makes the code idiomatic and preferable [1], [32]. For instance, professional programmers often prefer compound operators (e.g., `x += 4`) over their verbose counterparts (`x = x + 4`).

Code structure can be analyzed by examining specific parts of the code in isolation, without requiring an understanding of the broader problem context. For example, when checking a condition and its negation, using `if` and `else` is generally preferred over using two if statements of `if(a)` and `if(!a)`. Code structure is distinct from quality aspects that do not affect code execution, such as comments, and from broader aspects, like object-oriented design, which involve multiple functions or classes. Prior studies have used other terms such as *semantic style* [1] and *programming patterns and anti-patterns* [33] to refer to the same aspect. Table I shows some examples of well-structured code and code structure violations.

Despite the importance of writing well-structured code, this task is challenging for students [11], [33]. Code structure violations do not result in compiler or syntax errors, making it difficult for students to recognize these issues on their own. Additionally, since students are often graded based solely on code functionality, they often don't receive code structure-related feedback. Thus, they may lack the motivation to spend extra time reviewing their work [33], [34]. Some studies even suggested that student violations may stem from deeper issues, such as misunderstandings or knowledge gaps related to language constructs [1], [35]. Therefore, it is not surprising

that student programs exhibit various code structure violations [11], [36], [37]. Unfortunately, students' code structure is unlikely to improve without targeted intervention, even as students gain more programming knowledge [1], [6]. Therefore, instruction on writing well-structured code is necessary. While many studies examined different approaches to teaching code structure, our field lacks an overview of the approaches and their effectiveness. Thus, we conducted this review study.

*B. Review Studies on Code Quality in Educational Contexts*

There has been limited research reviewing the literature on code quality and style in educational settings. A recent systematic mapping study by Keuning et al. [38] provides an overview of studies conducted over the past decades, identifying their topics, research methods, and languages used. Building on this study, we narrowed our focus to a specific subset of the 195 papers they identified while also incorporating more recent studies published in 2023.

Starke and Michaeli's poster [39] reviewed studies on code quality in K-12 education, where block-based programming environments are commonly used. These environments are not immune to code smells, which can disrupt students' understanding of programming concepts. Their work identifies activities like refactoring, linter usage, and code reviews as promising strategies to teach K-12 students about code quality. However, we are not aware of any full systematic review.

### III. METHODOLOGY

We conducted a systematic literature review to identify instructional approaches used to teach function-level code structure, examine how their effectiveness was evaluated, and highlight gaps requiring further investigation. We built on Keuning et al.'s [38] systematic mapping study by starting with their dataset of 195 papers on code quality published by the end of 2022. We used their search string to search Scopus, IEEE, and ACM databases to identify relevant papers published in 2023. Since our focus is narrower—code structure rather than code quality—we introduced further exclusion criteria. To answer our RQs, we developed codebooks. The following sub-sections detail these steps.

*A. Keuning et. al's Systematic Mapping Study*

The systematic mapping study by Keuning et al. [38] provided a high-level overview of research on code quality in educational settings. The authors developed a search string to conduct a search in three databases: Scopus, ACM, and IEEE—main repositories of peer-reviewed research that support advanced search functionality—to collect papers up until the end of 2022. The search covered papers' titles, abstracts, and keywords. To identify the relevant papers, they applied the following inclusion and exclusion criteria.

– Included English-language studies that focused on code quality within an educational context, where a significant portion of the paper was dedicated to this topic.
– Excluded papers unrelated to education, such as studies examining smells in software by professional developers.

> (programm* OR code OR coding OR software)
> AND ("code quality" OR "software quality" OR "design quality" OR refactoring OR "static analysis" OR "software metrics" OR smell OR readability OR "code style" OR "coding style" OR "programming style")
> AND (student OR teach* OR educat* OR curriculum OR novice)

Fig. 1: The search query utilized by Keuning et al. [38] for the systematic mapping. We used the same query but applied further exclusion criteria to the identified papers.

– Excluded papers shorter than four pages, papers focused on domain-specific languages like SQL, studies that discussed interventions that did not primarily focus on code quality, and the ones that involved student participants without an educational purpose.

After applying the inclusion and exclusion criteria, 168 papers were selected for inclusion. After snowballing, 27 more papers were added to the dataset.

*B. Our Dataset and Additional Exclusion Criteria*

We used Keuning et al.'s [38] search string to collect papers from 2023. This search yielded an additional 90 papers from IEEE, 84 papers from ACM, and 81 other Scopus papers. Because our scope was narrower than Keuning et al.'s [38], we defined additional criteria to exclude papers that solely focused on the following without discussing code structure:

– Aspects of code quality that do not affect code's execution, such as comments, the naming of variables (e.g., [40]), or formatting issues like indentation and whitespace (e.g., [41]).
– Code quality topics involving multiple functions, classes, OOP code structure, or design smells, that did not address function-level code structure (e.g., [42]).
– Quality metrics like cohesion or complexity without discussing individual code structure topics (e.g., [43]).
– Language-specific structures (e.g., built-in Python functions [44]) that can not used across multiple languages.
– Programming templates (e.g., [31]) and approaches to the problem (e.g., [45]).

We applied our exclusion criteria to the 195 papers included in the mapping study [38] and to the additional 255 papers from 2023. To apply the exclusion criteria, we initially read the paper's title, followed by its abstract. If we couldn't decide at this stage, we proceeded to skim through the paper to locate the section related to code quality and assess whether it covered any function-level code structure topics. After examining all papers, 67 papers were included. Next, we performed a one-step snowballing on the papers that were not part of Keuning et al.'s dataset by looking at their references (we also looked for papers that cited the newly-included papers, but the 2023 papers had been published for less than a year when we conducted this analysis in December of 2023). Snowballing resulted in the inclusion of three additional papers. In the next step, we needed to identify studies on the *instruction* of code

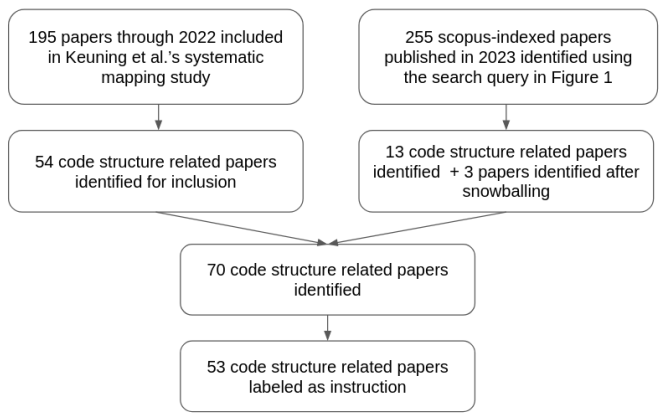

Fig. 2: Flow diagram for systematically extracting the papers. We used the studies identified by Keuning et al. [38] up to the end of 2022 and searched for 2023 papers using their query. We then applied our exclusion criteria to all studies.

structure, as some studies in our dataset focused on other topics, such as the *assessment* of code structure. Deciding on which study to include as *instructional* was challenging. For instance, some studies discussed using automated tools to identify code structure violations in student code as a means of assessing their skills without addressing the tools' instructional applications. When we were uncertain about such studies, we classified them as *instructional* because such tools could be potentially useful for educational purposes and help students identify and correct violations in their programs. Similarly, some papers provided resources or materials (e.g., a catalog of code structure violations) that could be potentially useful for educational purposes. These catalogs were often derived from running code analyzers on students' submissions, and the authors did not showcase how instructors could integrate the catalogs into their teaching and did not evaluate them as instructional tools (e.g., [13]). We labeled all papers that either explicitly mentioned instruction as their objective or could reasonably be applied in instructional contexts as *instructional*. This criterion resulted in the inclusion of 53 papers for this study. Figure 2 details this process.

### C. Developing the Codebook for Our RQs

To develop the codebook, we began by brainstorming topics of interest (e.g., instruction, evaluation) and an initial set of labels to apply to each paper (e.g., *discussed an instructional approach* vs. *potentially useful for educational purposes*). Note that we use the term *label* rather than *code* to avoid confusion with our programming context. With the initial topics in mind, the first two authors independently labeled five randomly selected papers from the dataset. Then, they conferred to resolve disagreements and discuss any new labels they had generated (e.g., to describe how the instruction was evaluated). After refining the codebook, they independently labeled five more randomly selected papers. After this round, they determined that the codebook was stable: they resolved disagreements about the application of the labels, but did not need to add or remove labels from the codebook. After refining the criteria for applying each label, they divided the full set of papers and labeled each half independently (including re-coding the initial set of five to align with the final codebook). The two authors discussed all cases where they were uncertain about a label. After the independent labeling was complete, the first author reviewed all labels, resolving any disagreements through discussion.

When labeling the papers, the authors made some notes about each study, including the type of artifact it involved (e.g., catalog, rubric). Since these notes included valuable information about the type of artifact the studies discussed, the first author developed a set of labels to categorize the type of artifacts. Subsequently, both the first and second authors applied these labels to all papers individually and resolved any disagreements.

We now discuss the labels (in *italics*). Each paper either:
- *Discussed an instructional approach* in detail, with the clear objective of teaching students, or
- Did not discuss explicit instructional applications, but presented a tool or resource that could be *potentially useful for educational purposes*.

Each study was classified for its instructional approach into only one of the following:
- A supportive *tool* was discussed only within its context of use and was used without other teaching materials for code structure.
- Other *teaching materials* were developed or discussed either solely or in conjunction with tools (e.g., a rubric).
- *Integrating code structure into the curriculum*, discussing a holistic approach to course design to support code structure.

We also labeled each paper with at least one of the following artifacts the study described.
- *Custom-made* educational tool.
- *Professional* tool.
- *Rubric* for feedback or grading of code structure.
- *Catalog* of code structure violations or misconceptions.
- *Learning resources*, including interventions such as refactoring lessons, guidelines, activities, or exercises for code structure.

For each paper that *discussed an instructional approach*, we labeled the stated goal(s) as at least one of:
- *Improving students' code structure*, code quality, increasing code structure/quality scores, or reducing defects in student programs.
- *Improving students' awareness* of code structure topics and their violations or increasing students' motivation and engagement with code structure.
- *Other* goals mentioned, such as improving students' comprehension of code.

Each paper that *discussed an instructional approach* either:
- *Measured the effectiveness* of the approach (for some impact on students), or

- *Did not measure the effectiveness* of the approach for impacts on students. Some papers with this label measured other types of effectiveness (e.g., performance of a tool in identifying defects).

Each paper that *measured the effectiveness* of their approach was labeled with at least one of the following evaluation types:

- Student *learning* of code structure measured by comparing their knowledge before versus after the intervention (i.e., without using the tool or instructional material).
- Student *code structure performance* was measured. In this approach, the researchers compared student performance before and when using the intervention or the performance of students who used the intervention with those who did not. The key difference from measuring learning is that performance is measured while students have access to the target tool or instructional material. To assess performance, the researchers may have used factors such as overall grades, code quality scores, frequency of violations, mean of violations, and other relevant metrics.
- Student interaction or *behavior* examined by analyzing the number of times they engaged with the intervention (e.g., the number of times they requested feedback).
- Student *self-report* responses to survey questions were examined, including their perceptions of the intervention's usability, self-reported learning, and other related aspects.
- *Other* methods were used.

## IV. FINDINGS

Overall, 53 studies were labeled as *instructional*. Of these, 37 studies *discussed an instructional approach*. In the remaining studies, we identified at least one artifact or resource that could *potentially be used for educational purposes* even though it was neither specifically designed nor discussed in detail for instruction.

### A. RQ1: Among Instructional Approaches, Supportive Tools Were the Most Common

We found that 36 studies in our dataset primarily focused on supportive *tools*, 15 studies involved *teaching materials*, and two studies detailed the researchers' experience of *integrating code structure into the curriculum* (see Table II for citations).

TABLE II: Instructional approaches from studies on code structure instruction. The numbers in parentheses indicate the number of studies discussing each approach.

| Instructional Approach | Discussed the Approach | Potentially Useful |
|---|---|---|
| Focused on supportive *Tool*s and their usage (36) | [15]–[17], [26], [27], [46]–[64] | [11], [23], [36], [37], [65]–[72] |
| Involved non-tool *teaching materials* (15) | [13], [18]–[22], [24], [73]–[76] | [1], [77]–[79] |
| *Integrating code structure into the curriculum* (2) | [80], [81] | |

We organize our findings for RQ1 using these three instructional approaches. However, since some papers discussed multiple instructional artifacts (e.g., a professional tool and a catalog), we labeled papers for all relevant artifacts. We discuss these artifacts within the approaches they are most relevant to, and for each artifact, we provide example studies to showcase it. Within each approach, we first discuss studies that *discussed an instructional approach* and then studies *potentially useful for educational purposes*. Table III provides a summary of the artifacts discussed in each study, including citations.

*1) Instructional Approaches Centering on Supportive Tools:* Of the studies involving supportive *tools*, 24 *discussed an instructional approach*, such as refactoring tools, tutoring systems, and code analyzers. We found that the authors' primary goal for using or discussing a supportive *tool* was *improving students' code structure*. Additionally, six studies, aimed to *increase students' awareness* of various aspects of code structure and quality [16], [17], [48], [57], [60], [64], while two studies also mentioned supporting instructors in providing feedback and grading [46], [52].

Supportive tools can be broadly categorized as *custom-made* or *professional* tools. In most cases, these *custom-made* tools make use of existing professional tools, components, or API's.

*a) Custom-made Educational Tools:* Many studies in our dataset discussed using *custom-made* tools specifically for instructing students. Some educational tools were completely custom-built, including the violations to be identified, hint or feedback messages, and the interface (e.g. [57], [64], [76]). However, others leveraged *professional* ones (e.g., [16], [17], [58]). While leveraging professional tools allows researchers to utilize existing detectors and perform a broader range of checks on student code, they are designed for professional developers and are not tailored to educational needs. Thus, researchers often configure the ruleset to retain only those relevant to students [17], [61]. Some researchers also modified the feedback messages to enhance student comprehension [27], [48], [61], and some integrated custom checkers into their tools [27], [58] to address novice-specific violations that professional tools may overlook [11], [23].

Among studies that discussed *custom-made* tools, two studies discussed a refactoring tool that could automatically perform refactorings on student violations [57], [76]. For example, Techapalokul and Tilevich [57] defined four refactoring rules for Scratch projects to address code duplication and variable scope. They developed a refactoring tool capable of determining the optimal order for applying these rules. The tool offers hints and refactoring suggestions to programmers, which they can implement manually or directly apply to the code with a single click.

In contrast to automated refactoring systems, the REFACTOR TUTOR [64] is a tutoring system designed to teach students how to refactor functional but poorly written Java programs, while enhancing students' awareness of code quality. The system allows students to request feedback on functionality (to ensure the code's behavior remains unchanged) and hints on how to improve code structure at any point during the

TABLE III: A summary of instructional artifacts from studies on code structure instruction. The numbers in parentheses indicate the number of studies discussing each artifact. The number of the studies does not sum to 53, as some discussed multiple artifacts. Custom-made and professional tools were the most common instructional artifacts.

| Artifact | Description of Artifact | Discussed an Instructional Approach | Potentially Useful for Education |
| --- | --- | --- | --- |
| Custom-made educational tools (31) | Researchers discussed an educational tool that can identify violations in student programs, provide feedback and/or hints for manual refactoring, perform the refactoring automatically, or teach students how to refactor. In some cases, researchers utilized existing tools and API under the hood. | [15]–[17], [26], [26], [27], [46]–[54], [56]–[59], [61]–[63], [76], [82] | [13], [23], [37], [65], [70], [78] |
| Professional tools (13) | Researchers utilized off-the-shelf professional tools for teaching code structure and code quality to students. | [13], [24], [60], [80], [81] | [11], [13], [36], [66], [68], [69], [71], [72] |
| Learning Resources (8) | The researchers offered materials such as a set of practice exercises, refactoring rules or lessons. | [13], [18]–[20], [24], [74]–[76] | |
| Rubrics (3) | The researchers designed rubrics to provide code structure-related feedback on students' coding assignments or exams. | [21], [22], [73] | |
| Catalogs (5) | The researchers developed a catalog of frequent violations in student programs, or the mistakes they make while refactoring. | | [1], [13], [77]–[79] |

refactoring process. However, it does not perform the refactorings for them. The REFACTOR TUTOR's hint messages are delivered progressively: first, a high-level overview, followed by more detailed guidance, and culminating in concrete code suggestions when necessary. The system's rules, developed with input from instructors, address code structure violations related to loops, conditionals, and expressions. Because the tool offers six targeted exercises, it cannot be applied to diverse student submissions, unlike automated refactorers [57], [76].

Most studies involving *custom-made* tools discussed educational code analyzers that automatically flag violations in student programs and provide feedback. These analyzers typically operate without instructor involvement, relying on predefined rules (e.g., [15], [56], [58]). However, some tools require instructor input [52], [62], [63]. For instance, SPT [62] uses an instructor-provided model solution to deliver strategies for arriving at that solution. The system also offers an instructor dashboard that provides reports of aggregated violations in student submissions. Some tools require instructors to define the violations the tool should identify, such as the JAVA CRITIQUER [52]. Both SPT and JAVA CRITIQUER allow instructors to confirm or remove violations from the final report shared with students.

*b) Using Existing Professional Tools in Educational Settings:* Some studies in our dataset discussed utilizing *professional analyzers* to engage students with code structure [24], [60], [80], [81]. For instance, the study by Alomar et al. [60] described a classroom experiment conducted over three semesters in which students reviewed software they had not written, using PMD to identify violations. For each defect identified, students were required to justify whether it should be fixed. The study analyzed which PMD flags students perceived as true or false positives and stated that integrating tools like PMD into software practice courses can enhance students' understanding of code structure and software quality concepts, while also fostering a culture of quality improvement. Another study [24] discussed using SONAR, a professional tool, to provide feedback to students on their violations. However, its effect on students has not been studied.

*c) Supportive Tools Not Specifically Discussed for Instruction but Potentially Useful for it:* Some studies in our dataset utilized tools to assess students' code structure violations and reported the prevalence of these violations without discussing the tool's use for instruction. Among these studies, some utilized a *custom-made* tool (e.g., [13], [23], [37], [70], [78], [79], and the rest used a *professional* tool [11], [13], [36], [66], [68], [69], [71], [72]. We see the potential for students to use these tools to locate violations in their code, while instructors can use them to learn about common violations in student programs. This insight can enhance instructors' Pedagogical Content Knowledge by helping them understand students' challenges and misconceptions, enabling the design of targeted instruction and feedback [83].

Within studies that used *professional* tools, PMD was the most frequently used (e.g., [11], [36], [72]). Some other employed *professional* tools included SonarSource [36], CheckStyle [69], and Pylint [71]. Most of these studies did not discuss the feedback or information these tools provide. However, professional tools are typically designed to deliver feedback to individuals while programming, not to display classroom-level assessment data afterwards. Therefore, instructors may need to create custom scripts to aggregate violations across student submissions. This is also the case for some custom-made tools. Additionally, professional tools may not flag violations that are relevant for the learning goals of a course, and may need to be configured to avoid flagging violations that are not appropriate for students at a particular level.

*2) Teaching Materials:* In addition to tool-based instructional approaches, fifteen studies discussed other types of interventions and materials. Among these, eleven studies *discussed an instructional approach*. The main goal of all these studies was *improving students' code structure* with two studies also aiming at *improving students' awareness* [18], [24] and one study also having the goal of supporting instructors in providing feedback to students [21]. While these studies may have also involved *custom-made* or *professional* tools, they incorporated at least one of the following instructional artifacts: *rubrics*, *learning resources* such as an activity or exercise, and/or *catalogs* of violations or guidelines.

*a) Rubrics:* One supporting teaching artifact is the *rubric*, which can help instructors and students (self-)assess the quality of code, provide feedback, and incorporate code quality as part of the student grading [21], [22], [73]. Stegeman et al. [21], [73] developed a rubric to assess the quality of student code and to provide students with feedback. The rubric is based on a model of code quality derived from literature, including Clean Code [7], Refactoring [8] and Code Complete [84], and instructor's input [73]. Based on that model, the authors designed the rubric in iterations (using 'educational design research'), in which trial assessments were conducted, and findings were discussed with instructors. This process resulted in the first version of the rubric, which can be used by instructors, TAs, and students. The rubric includes criteria such as layout, expressions, flow, and idioms. Each criterion is broken down into specific, observable aspects that can be applied to student code. However, we did not find any research that has used this rubric with students.

The study by Whalley et al. [22] differs from this study in that they developed a rubric by analyzing code solutions to specific assignments, using the SOLO taxonomy and grounded theory to identify themes for the structures students used and grade students based on the structure. As a result, the rubric is tailored to those assignments. However, by following Whalley et al.'s approach, instructors can create similar rubrics for their own assignments, providing code structure-related feedback to students and incorporating code structure into their grading criteria.

*b) Learning Resources:* Some studies in our dataset provided resources such as refactoring lessons, exercises, rules and guidelines to help students practice and learn how to fix specific code structure violations or write well-structured code [13], [18]–[20], [64], [74]–[76]. For instance, Izu et al. [18] present a resource designed to teach students refactoring rules related to conditionals. The resource includes: 1) a one-page introduction to code quality with a rationale for its importance, 2) explanations of the rules with code examples for each, and 3) a set of three refactoring exercises. In a lab session of an introductory programming course, the researchers conducted an experiment in which students were provided with the introduction and refactoring rules (parts 1 and 2) and then asked to apply these rules to the refactoring exercises (part 3). The study examined student performance on the refactoring exercises and compared the number of tokens in student homework submissions two weeks after the intervention with corresponding submissions from the previous year, when students did not have access to the resource. They found that students who used the resource comparatively wrote shorter code. Another study [13] examined the effect of scaffolding in teaching a specific code structure topic: *directly return boolean expressions instead of using if and else*. When the researchers observed students frequently violating this topic, they added three problems with scaffolding for this topic to their learning environment. They found that scaffolding reduced the prevalence of the violation by 30%. However, the study does not provide details about the scaffolding intervention. This study was the only one examining the effect of scaffolding on teaching code structure.

A few studies in our dataset explored the use of code review activities to improve students' code structure [20], [24], [74]. The study by Hashiura et al. [24] explains their code review process in several steps: 1) Code writing: Students wrote programs based on the provided problem instructions. 2) Peer code review: Students were grouped, and each student's program was reviewed by another group using a review sheet that included aspects such as naming, algorithms, code structure, and more. 3) Correction: Programs were corrected based on the feedback provided during the review. 4) Final deliverables: Students presented the finalized versions of their programs after incorporating the corrections. In this study, the authors found that conducting peer code reviews decreased the frequency of violations, but the reduction varied based on the topic. The study by Andrade et al. [74] did not include the code writing step, and students reviewed code written by students in previous terms. The authors found that students' accuracy in identifying errors depended on the topic, but in general, their feedback was 50% or more similar to the feedback from instructors, and most students were able to provide useful feedback to their peers. The study by Gaber et al. [20] differs from the previous two studies, as in this study, students were asked to review their own code and provide reports of issues, including code structure violations, bugs, and missing features that they could not fix. The authors found that students reported missing features and bugs quite accurately but were less aware of code structure violations (less than 30% of violations were reported).

*c) Teaching Materials Not Specifically Designed for Instruction but Potentially Useful for it:* Some studies labeled as *potentially useful for educational purposes* provided approaches beyond supportive tools [1], [13], [78], [79]. These studies offered a *catalog* of violations or misconceptions. We provide example studies to explain how we think *catalogs* can be used for teaching code structure.

Some other studies in our dataset also involved sets of code smells or violations collected by analyzing student programs using various tools. However, we only labeled papers as *catalog* when the set of violations was explicitly presented as some kind of catalog and mentioned as part of the authors' contribution [1], [13], [77]–[79]. While no study empirically explored the use of these catalogs in an educational setting,

one potential application is to use them as a reference list of violations for students to identify while reviewing their own or peers' code. An example of a study offering a *catalog* of violations is Effenberger and Pelánek's work [13], which presented a catalog of 32 code quality defects derived from analyzing numerous solutions to various Python programming problems. The catalog categorizes defects into topics such as loops, conditionals, expressions, variables, and functions, providing an example of each defect along with its solution. The study by Oliveira et al. [79] differs from the other studies that offered a *catalog*, as it presents a *catalog* of refactoring misconceptions—mistakes students make while refactoring code. The authors advise instructors to address code refactoring in their programming courses, teaching correct refactoring steps and possible misconceptions along the way.

*3) Integrating Code Structure into the Curriculum:* The two studies in this category discussed the authors' experience incorporating code quality into existing software development courses, detailing the steps taken, challenges encountered, and lessons learned [80], [81]. Both studies aimed to *improve students' code structure*, and the study by Sripada et al. [81] also aimed for students to practice code-reading.

The study by Vasileva et al. [80] described three iterations of a software practice course aimed at incorporating code quality, with each iteration designed to address shortcomings identified in the previous iteration. In the first iteration, the instructors used PMD to assess students' code and identify areas for improvement. Additionally, they determined which PMD flags were relevant for students and set appropriate thresholds. Running PMD on student submissions revealed that student programs suffered from various violations. In the second iteration, they put more emphasis on teaching code quality throughout the software development process. Clean Code [85] and Refactoring [8] principles were introduced, with tutorials on static analysis and refactoring using Eclipse and PMD. Then, students were asked to develop high-quality software. However, students' software did not show sufficient quality improvement. The authors stated that addressing code quality only at the end of the project left insufficient time for effective refactoring. In response, the third iteration aimed to highlight the challenges of removing code smells late in the project, motivating students to write clean code from the start. To reinforce motivation, instructors organized a competition where the team with the best code quality could win a prize. In addition, instructors measured the projects' quality with students and discussed potential errors and violations, as well as how they could identify bad practices and improve them in the earlier stages. Moreover, students were required to address the identified violations through refactoring. If they were unable to successfully implement the refactoring, they were required to write a report justifying the reason. In this iteration, students' overall code quality significantly improved. This study [80] is the only study in our dataset that demonstrates how instructors can enhance their Pedagogical Content Knowledge by analyzing students' code structures. The study highlights that integrating code quality into the curriculum is challenging and time-intensive. However, it can be implemented gradually over multiple semesters, allowing instructors to refine their approach.

The study by Sripada et al. [81] describes how they integrated code quality into a Software Engineering course through continuous code review activities. The course involved four phases. In phase 1, students learned about coding standards and reviewed their own code for standard-related issues. In phase 2, students worked in teams to develop software and conducted code reviews of other teams' work, both manually and using tools. In phase 3, students were introduced to code smells, learning how to identify and refactor them. They reviewed other teams' code and applied refactoring to their own projects based on peer feedback. In the final phase, students delivered their completed projects. The students reported that the continuous code review process had improved their code comprehension skills [81]. Additionally, the authors provided suggestions for instructors who want to use code review in their courses. This included providing checklists of bugs and violations for students, limiting the number of lines of code to review at a time, and using external review platforms such as Gerrit instead of lightweight integrated tools [81].

*B. RQ2: Among Methods for Examining the Impact of an Instructional Approach, Comparing Students' Performance was the Most Common*

Of 37 studies that *discussed an instructional approach*, only 20 *measured the effectiveness* of their approach for impact on students. Among these, 13 studies involved supportive *tools*, two involved *integrating code structure into curriculum*, and five involved *teaching materials*. Additionally, nine studies employed multiple evaluation methods [16]–[18], [20], [27], [57], [61], [74], [81]. We found that researchers used various approaches, with measuring student's *code structure performance* being the most common approach (16 studies). These studies took one of the following approaches: (1) comparing the violations in student programs before and when using the intervention [13], [18], [26], [58], [61], [81]; (2) comparing violations in programs written by students who used the intervention with those written by students from previous years who did not have access to the intervention [16], [17], [20], [24], [27], [56], [62], [63], [80]; and (3) comparing violations in programs from students who used the intervention versus those in a control group within the same cohort [57]. The second common approach was to ask students to *self-report* their perceptions or performance (nine studies) [15], [17], [20], [26], [27], [50], [60], [61], [81]. We found that researchers used *self-report* in different ways. For example, one study asked students to name three key lessons they learned from using the intervention [17]. Another had students report how often the feedback led them to change their responses [15]. Some studies gathered student responses about the usefulness of the intervention and/or the understandability of its feedback messages [20], [50], [57], [60], [61], [81]. Three studies investigated students' *behavior*, focusing on how students interacted with the intervention [16], [26], [27]. While these studies

reported on students' code structure performance, they also examined log data and performance metrics to more accurately understand how the intervention was used. For example, they examined how often students requested feedback from the tool [26], [27], the number of attempts to resolve a single violation [16], [37], and how students acted on feedback to assess whether the messages were understandable [26], [37].

Four studies in our dataset utilized *other* evaluation methods [20], [60], [74], [81]. One study [60] asked students to use PMD to identify violations in the software they had not written and determine which of the PMD flags they considered as true or false positives. Three studies examined the effect of code review activities by asking students to identify violations in their peers' code [74], [81] or their own code [20]. The researchers analyzed the identified violations and the extent to which students could recognize these violations. The two studies that included peer code reviews [74], [81] also examined the feedback students provided to one another. One study conducted sentiment and subjectivity analyses of the feedback [81], while the other [74] analyzed the similarity between instructors' and students' feedback on the same violations. Additionally, instructors were asked to classify students' feedback as either useful or not [74]. Only one study in our dataset investigated the impact of the intervention on students' *learning* after they stopped using it [18]. In this study, researchers evaluated how teaching refactoring rules can affect students' code writing assignments completed two weeks after practicing with the rules [18].

Regarding the effectiveness of instructional approaches, all studies in our dataset that *measured the effectiveness* of their approach with students reported evidence of effectiveness in at least one of the methods used. However, the diversity of approaches prevents direct comparison between different types of interventions. Additionally, we found that nearly all studies involved classroom experiments (except for [86]), and we found no randomized controlled trials on the effectiveness of these interventions.

We provide our codebook, the complete labeling of all papers and a summary of the approaches each study used to examine the effectiveness of their approach and its findings as supplemental material.[1]

## V. Discussion

### A. Instructional Approaches, Their Benefits, and Limitations

Over half of the studies in our dataset used supportive *tools* to assess violations in student code, provide feedback to students, and offer insights to instructors. While scalable, tools often offer limited control over which violations are detected and how feedback is presented. Thus, they may be less suitable for intro-level students. However, instructors can utilize tools that support rule configuration. In addition, we observed that many studies in this category did not discuss the tool's feedback messages (e.g., [50], [56]). While some studies (e.g., [50], [61], [64]) surveyed students about the clarity of

[1]https://doi.org/10.5281/zenodo.14872900

feedback messages or analyzed how students responded to feedback [26], [37], many did not. Consequently, it is unclear whether the feedback messages provided by these tools are adequate for students. Additionally, few studies have explored the optimal level of feedback for students [26], [33]. Thus, we encourage further research into what kind of feedback and support students require to address different violations.

Approaches involving *teaching materials* were less common. Among these, researchers discussed non-tool artifacts, including learning resources such as refactoring lessons and activities, as well as supporting rubrics for manual grading and feedback. These approaches are often more time-consuming than using tools, but they offer more flexibility. For example, if an instructor prefers to micro-insert code structure (adding quick and low-effort activities or discussions related to code structure) into a CS1 course, similar to the study by Gaber et al. [20], they can ask students to also submit a report of the violations they can identify in their code. Similarly, instructors can provide students with a list of violations and ask them to review each other's programs for those violations [24]. Another approach is to teach aspects of well-structured code and refactoring alongside core programming concepts. For example, when teaching `if` and `else` structures, instructors can also discuss the importance of avoiding repeated code or logic within the branches and demonstrate how to refactor such code. Instructors can use catalogs of defects to decide on which violations to prioritize. In some studies, researchers discussed exercises, scaffolded tasks, and lessons to directly teach students how to refactor specific violations [18], [64]. We found that the study by Effenberger et al. [13] was the only one in our dataset on the effect of scaffolding in teaching code structure. Thus, we encourage future studies in this area. Potential scaffolding strategies could include fill-in-the-blank tasks, where students complete partially written code by filling in the relevant structure, and Parson problems with distractors, where students choose the correct structure to use. We recommend future design of such tasks using the catalogs we identified.

Beyond the instructional approach categories used in this study, we can also subdivide approaches in (1) providing feedback on student violations (e.g., using code analyzers, rubrics), and (2) directly teaching code structure (e.g., through refactoring lessons [18], [87] or activities [26]), which requires designing targeted tasks and materials. The key difference is that the first approach targets only students who make mistakes, while the second engages all students in practicing code structure. While the first approach may seem sufficient, existing tools have limitations in identifying important violations. Moreover, students who do not violate a topic do not necessarily master it [23] and may violate it in another task.

### B. A Variety of Methods To Examine the Effectiveness of Instructional Approaches

We found that comparing student performance in code structure was the most commonly used evaluation method. This involved analyzing the frequency of errors in specific topics

(e.g.,[18], [26]) or using cumulative metrics such as mean or median error counts (e.g., [16]). While both methods are valid, the former offers more detailed insights into the effectiveness of an approach in addressing specific code structure violations, especially since prior research suggests that a single instructional approach may not be equally effective for different violations [6], [16], [24]. Moreover, even when an instructional approach leads to improved student performance, it does not necessarily indicate that students can independently identify and resolve issues. Further research is needed to determine whether students continue addressing code violations after discontinuing the intervention.

We also found that almost all research in our field include classroom studies, which means there is a lack of controlled lab studies on the effectiveness of different approaches. While classroom studies have high ecological validity, randomized control lab studies often have higher internal validity and allow researchers to draw stronger causal inferences about the effectiveness of different approaches. Finally, we observed that researchers employed diverse methods to evaluate their interventions, involving varying numbers of participants from different levels. This variability makes direct comparisons between interventions challenging.

*C. Advice to Researchers and Instructors*

We encourage further research into approaches beyond tools and their impact on students' learning of code quality and programming in general. We particularly suggest exploring how catalogs can guide the design of teaching materials—such as refactoring lessons, scaffolded code-writing tasks, and activities like code review and refactoring. Manual code reviews and code refactoring require students to read, understand, and analyze their own as well as others' code to identify violations and implement improvements. All these tasks are essential for developing professional programming skills. Students' over-reliance on tools may impede the development of these skills. One potential approach is to provide students with code snippets containing violations and ask them to independently identify and fix the issues. An automated solution is demonstrated by Keuning et al. [26] in the REFACTOR TUTOR, which offers students progressive hints after they attempt to identify violations on their own.

Writing well-structured code requires multiple skills [33], [88], such as understanding and applying appropriate structures (e.g., we can directly return boolean expressions or variables). Some studies suggest that student violations in certain topics arise from underlying knowledge gaps [1], [15]. Therefore, further research is necessary to identify these gaps and design targeted interventions to address them. Identifying these gaps is also essential for improving the design of tools' feedback, as students may require additional explanation. Moreover, experimental studies evaluating the effectiveness of feedback provided by tools remain relatively rare. Yet, offering the right level of feedback is critical—minimal feedback may be unclear, while overly detailed feedback can be unnecessary and increase cognitive load.

VI. LIMITATIONS

Like all literature reviews, ours is limited by our defined scope (our definition of *code structure,* which excludes class-level issues) and our search methods (which may have missed relevant papers even within our target databases, and excluded all papers outside of them). Labeling papers as *instruction* or not proved to be quite challenging, particularly for the label *potentially be used for educational purposes*, as we could envision some instructional potential in all the papers in our dataset. Consequently, we established specific criteria to follow. For instance, determining if a paper *measured the effectiveness* of their intervention with students was not a clear cut. While some studies examined their tool by running it on previously submitted students' solutions and reported on the violations identified, this type of evaluation does not show whether and how the tool impacts students. Therefore, we only labeled papers as *measured the effectiveness* of their intervention if students interacted with the intervention. Although the labeling required interpretation, we hope our definitions of the labels and the fully labeled dataset as online resource provides the necessary transparency for others to interpret our results.

VII. CONCLUSION

We presented a systematic literature review to identify instructional approaches that have been described in prior studies to teach computing students about writing well-structured code. We built upon an existing mapping study of code quality in education, using a subset of their paper dataset as a starting point and extending with more recent papers up to and including 2023. We investigated which approaches have been tried, what goals the authors had with the approach, if and how they tested the effect of the approach.

We found 53 papers that discuss an instructional approach to teach code structure, or an artifact that we think can potentially be used for instruction. Few of them involved teaching materials, such as specific exercises, rubrics, and activities. Many approaches revolve around digital tools that point students to problems in their code, or help them with fixing such problems. Notably, only a few studies emphasized manual review of others' code—a practice that could become increasingly important as Generative AI tools are now capable of producing code. Among the 53 papers, 37 studies discussed an approach specifically to education with the goals of helping students improve their code structure or raising their awareness of code quality. However, 17 of these studies did not evaluate the effectiveness of their approaches. Even among those that did, there is limited evidence to suggest that the effects of the interventions persist over time or transfer to related tasks.

Our contribution provides an overview of instructional approaches for teaching novices to improve code structure. We advocate for the development of more diverse instructional materials and encourage experiments to gather further evidence of their effectiveness for student learning and performance.

VIII. ACKNOWLEDGEMENTS

This work was supported by the National Science Foundation through the award SHF 1948519.